\documentclass[a4paper,fleqn,usenatbib]{mnras}
\usepackage[T1]{fontenc}
\usepackage{ae,aecompl}
\usepackage{enumerate}

\usepackage{graphicx}   
\usepackage{amsmath}    
\usepackage{amssymb}    
\usepackage{ulem}
\usepackage{color}

\title[Shang et al.]{Investigating multi-frequency pulse profiles of PSRs B0329$+$54 and B1642$-$03 in an inverse Compton scattering (ICS) model}

\author[L. H. Shang et al.]{
L. H. Shang$^{1,2}$,
J. G. Lu$^{3,4}$,
Y. J. Du$^{5}$,
L. F. Hao$^{6}$,
D. Li$^{2}$,
K. J. Lee$^{7}$,%
Bin Li$^{8,9}$,
L. X. Li$^{7}$,%
\newauthor{
G. J. Qiao$^{3}$\thanks{E-mail:\href{mailto:gjn@pku.edu.cn}{gjn@pku.edu.cn}},
Z. Q. Shen$^{8,9}$,%
D. H. Wang$^{1}$,
M. Wang$^{6}$,
X. J. Wu$^{3}$,
Y. J. Wu,$^{8,9}$,
}
\newauthor{
R. X. Xu$^{3,4,7}$,%
Y. L. Yue$^{2}$,
Z. Yan$^{8,9}$,
Q. J. Zhi$^{1}$\thanks{E-mail:\href{mailto:qjzhi@gznu.edu.cn}{qjzhi@gznu.edu.cn}}},
R. B. Zhao$^{8,9}$,
R. S. Zhao$^{8,9}$
\\
$^{1}${School of Physics and Electronic Science, Guizhou Normal University, No. 116, Baoshan Road, Yunyan Distric, Guiyang 550001, China}\\
$^{2}${National Astronomical Observatories, Chinese Academy of Sciences, 20A Datun Road, Chaoyang District, Beijing 100012, China}\\
$^{3}${School of Physics, Peking University, No. 5, Yiheyuan Road, Haidian District, Beijing 100871, China}\\
$^{4}${State Key Laboratory of Nuclear Science and Technology, Peking University, No. 5, Yiheyuan Road, Haidian District, Beijing 100871, China}\\
$^{5}${Qian Xuesen Laboratory of Space Technology, NO. 104, Youyi Road, Haidian District, Beijing 100094, China}\\
$^{6}${Yunnan Astronomical Observatory, Chinese Academy of Sciences, No. 396, Yangfangwang, Guandu District, Kunming 650011, China}\\
$^{7}${Kavil Institute for Astronomy and Astrophysics, Peking University, No. 5, Yiheyuan Road, Haidian District, Beijing 100871, China}\\
$^{8}${Shanghai Astronomical Observatory, Chinese Academy of Sciences, No. 80, Nandan road, Shanghai 200030, China}\\
$^{9}${Key Laboratory of Radio Astronomy, Chinese Academy of Sciences, No. 80, Nandan road, Shanghai 200030, China}\\
}

\date{Accepted XXX. Received YYY; in original form ZZZ}

\pubyear{2016}
\begin{document}
\label{firstpage}
\pagerange{\pageref{firstpage}--\pageref{lastpage}}
\maketitle

\begin{abstract}
The emission geometries, e.g. the emission region height, the beam
shape, and radius-to-frequency mapping, are important predictions
of pulsar radiation model. The multi-band radio observations carry
such valuable information.
In this paper, we study two bright pulsars, (PSRs B0329$+$54 and
B1642$-$03) and observe them in high frequency (2.5\,\rm{GHz},
5\,\rm{GHz}, and 8\,\rm{GHz}).
The newly acquired data together with historical archive provide
an atlas of multi-frequency profiles spanning from 100\,\rm{MHz}
to 10\,\rm{GHz}.
We study the frequency evolution of pulse profiles and the
radiation regions with the these data.
We firstly fit the pulse profiles with  Gaussian functions to
determine the phase of each component, and then calculate the
radiation altitudes of different emission components and the
radiation regions. We find that the inverse Compton scattering
(ICS) model can reproduce the radiation geometry of these two
pulsars. But for PSR B0329$+$54 the radiation can be generated in
either annular gap (AG) or core gap (CG), while the radiation of
PSR B1642$-$03 can only be generated in the CG. This difference is
caused by the inclination angle and the impact angle of these two
pulsars. The relation of beaming angle (the angle between the
radiation direction and the magnetic axis) and the radiation
altitudes versus frequency is also presented by modelling the
beam-frequency evolution in the ICS model. The multi-band pulse
profiles of these two pulsars can be described well by the ICS
model combined with the CG and AG.
%
\end{abstract}




\section{Introduction}
%

Since the first pulsar is discovered in 1967~\citep{Hewish68},
more than 2500 radio
pulsars\footnote{\href{http://www.atnf.csiro.au/research/pulsar/psrcat/}{http://www.atnf.csiro.au/research/pulsar/psrcat/}}~\citep{Manchester05},
and 200 gamma-ray
pulsars\footnote{\href{https://confluence.slac.stanford.edu/display/GLAMCOG/Public+List+of+LAT-Detected+Gamma-Ray+Pulsars/}{https://confluence.slac.stanford.edu/display/GLAMCOG/\\Public+List+of+LAT-Detected+Gamma-Ray+Pulsars/}}
have been observed. A variety of models have been proposed to
explain their radiation processes~\citep{Beskin15}. However, up to
now, the pulsar radiation is still a question.
~\citet{Goldreich69} proposes that a rotating magnetic neutron
star holds a charge-separated magnetosphere. ~\citet{Sturrock71}
quickly realized that a photon with energy larger than $2 m_e c^2$
can generate electron-positron pairs to produce radio emission by
the curvature radiation (CR). Subsequently,~\citet{RN75} develops
Sturrock's model. They propose the first accelerating model which
is called ``RS model", which presents an accelerating regions
(core gap model or CG model) above the polar cap of pulsar to
accelerate the particles. These particles move out along the
curved open magnetic field lines to produce radiation. The CG
model is used widely in pulsar research.

In recent years,~\citet{Q04} presents other accelerating region model,
i.e., annular gap (AG) model, to understand the particles
accelerating. ~\citet{Q04} suggests that there is an AG
accelerating region above the polar cap of pulsars around the CG.
The boundary of CG and AG is the critical field line
(CFL)~\citep{Holloway75}. In these two regions, the radio emission
is believed to be generated by inverse Compton scattering (ICS)
process between the secondary relativistic particles and low
frequency waves~\citep{Qiao88a,Qiao88b,Zhang96,QO98}. The ICS
model can explain the pulse profiles with peaks number from one to
five~\citep{QO98,ZG07}, the widening or narrowing profiles with
frequency increasing, the ``mode changing"
behavior~\citep{Qiao96,Zhang97a,Zhang97b}, and the
polarization~\citep{Xu97,Xu00,Xu01}. Besides, this model combined
with CG and AG can explain many observation phenomena at high
energy
bands~\citep{Q04,Lee06a,Lee06b,Lee09,Lee10,Du10,Du11,Du12,Du13,Du15}.

Multi-band observation of pulsar is an effective tool for testing
and constraining theoretical model. By analyzing the observed
multi-frequency pulse profiles, the location, geometry, and
spectrum of pulsar emission can be determined thus to constrain
theoretical model. On the basis of the empirical classification
for the observational properties of radio
pulsars,~\citep{RN83,RN93} proposes that the radio beam consists
of two hollow cones and a quasi-axial core normally. The debate
between the ``core-cone" beam models and other models (such as the
``patchy" beam, fan-beam, microbeam pattern et al.) as well as
multi-frequency observed data have been persisted by many authors
(e.g.
~\citealt{Lyne88,Kramer94,MR95,Gil96,Melrose99,Mitra99,Gang01,Han01,Kijak02,Manchester05,Karastergiou07,Maciesiak11a,Maciesiak11b,Maciesiak12,Beskin12,Wang12,Fonseca14,Wang14,Cerutti16,Dyks16,Pierbattista16,Teixeira2016}).
Besides, some theoretical perspectives (such as: the radio
radiation beam is a hollow beam without a core cone; the pulse
profiles at lower frequency are wider than those at higher
frequencies) have been challenged by a large number of
observations. The great variety of the pulse profiles as well as
their frequency evolution indicate the high complexity of the
pulsar beam patterns.

The multi-frequency pulse profiles characteristics of PSRs
B0329$+$54 and B1642$-$03 challenge some theoretical models (e.g.
the CR model). PSR B0329$+$54 is one of fairly bright pulsar in
the northern sky. It has a spin period of
P=0.715\,\rm{s}~\citep{2004MNRAS.353.1311H}. The pulse profiles of
PSR B0329$+$54 have obvious central peak~\citep{RN83,RN93}, and
the form of mean pulse profiles shows five ~\citep{KR94} or even
nine components ~\citep{Gang01,GA03,Chen11}. The pulse profiles of
this pulsar show narrower profiles at higher frequencies. PSR
B1642$-$03 has a spin period of
P=0.388\,\rm{s}~\citep{Manchester05}. The pulse profiles of PSR
B1642$-$03 consists of a core and an inner cone component, and the
pulse profiles at higher frequencies are wider than those at lower
frequencies~\citep{RN83,RN93,KR94}. These observed profiles
properties of this two pulsars can hardly be understood in some
theoretical models. Furthermore, as far as we know, no attempt has
been made to understand a broad-band pulse profile evolution
within the framework of most models. Therefore, it is of great
significance to do more multi-band observations of pules profiles
to test and constrain pulsar radiation mechanism model.

The major aim of this paper is to investigate the pulse profiles evolution over a two order of magnitude
frequency range, and study the pulsar radio radiation model, particularly the ICS model.
In~\S~\ref{Sect.2}, we will present the observed data of these
sources collected from the on-line database and the new
observations by the Kunming 40 meters Telescope, Yunnan
Astronomical Observatory, Chinese Academy of Sciences (CAS) and
Tianma 65 meters Telescope, Shanghai Astronomical Observatory,
CAS.
In~\S~\ref{Sect.3}, we separate the emission components of the
integrated pulse profiles of PSRs B1642$-$03 and B0329$+$54 by the
Gaussian fitting, calculate their radiation altitudes and present
the radio radiation regions of these two sources.
Conclusion and discussion are made in~\S~\ref{Sect.4}.
\begin{table*} 
\centering
\small
\caption{Observational information\label{1}}
\begin{tabular}{cccccccc}
\hline
Center frequency\,(MHz)& Telescope  & Bins & Bandwidth\,(MHz)& System temperature\,(K)  &SEFD$^a$\,(Jy)& Duration\,(s)\\
\hline
 2256 & Kunming 40~m$^{(b)}$ & 512 & 251.5  & 80 & 350 & 13035.326\\
 5030 & Tianma 65~m $^{(c)}$ & 1024 & 800 & 20 & 25 & 1773.823  \\
8600 & Tianma 65~m $^{(c)}$ & 1024 & 800& 35 & 50& 3600  \\
\hline
\multicolumn{7}{l}{\scriptsize{Note: $^{a}$ System Equivalent Flux Density; $^{(b)}$:~\citep{HO10}; $^{(c)}$:~\citep{YN15}.}}
\end{tabular}
\end{table*}

\section{OBSERVATION AND DATA REDUCTION}
\label{Sect.2}
The multi-band pulse profiles are very important for the detailed study of the frequency evolution of pulse profiles.
For PSR B0329$+$54, there are 10 profiles at frequencies 143, 408, 610, 925, 1642, 1410, 4750, 4850, 8500 and 10550\,\rm{MHz}~\citep{PA15,GD98,VH97,KR94}, while for PSR B1642$-$03, there are 8 profiles at frequencies 610, 925, 1408, 1410, 1642, 4850, 4750 and 10550\,\rm{MHz}~\citep{SS95,VH97,GD98} are obtained from the European Pulsar Network database\footnote{\href{http://www.epta.eu.org/epndb}{http://www.epta.eu.org/epndb}}.
In order to provide more profiles to constraint theoretical model.
We make observations at another three profiles at frequencies 2256, 5030 and 8600\rm{MHz}.
The pulse profiles at frequencies 5030 and 8600\rm{MHz} are obtained from the observations with the Tianma $65\,\rm{m}$ Telescope, Shanghai Astronomical Observatory, CAS, at longitude $121^\circ.1$ E and latitude $+30^\circ.9$ N.
The data were folded through Digital Backend System~\citep{YN15}.
The pulse profile at frequency 2256\,\rm{MHz} is obtained from the observation with the Kunming $40\,\rm{m}$ Telescope, Yunnan Astronomical Observatory, CAS, at longitude $102^\circ_.8$ E and latitude $25^\circ.0$ N~\citep{HO10}.
The data was folded through Pulsar Digital Filter Bank 3 with parameters provided by PSRCAT\footnote{\href{http://www.atnf.csiro.au/research/pulsar/psrcat/download.html}{http://www.atnf.csiro.au/research/pulsar/psrcat}}\,(version 1.51)~\citep{Manchester05}.
The radio-frequency interference (\,RFI\,) is removed manually
with the software package
PSRCHIVE\footnote{\href{http://psrchive.sourceforge.net}{http://psrchive.sourceforge.net}}~\citep{Hotan04,van
Straten12}.
The observed information is listed in Table\,\ref{1}.
\section{MULTI-band PROFILES ANALYSIS in the ICS MODEL}
\label{Sect.3}
The beam shape and the acceleration regions are very important to constrain pulsar emission mechanisms.
The emission beam contains two or three components, a center (core cone) and one or two nested cones (inner and outer cones).
We obtain the pulse profile when the emission beam sweeps across the line of sight.
The core, inner and outer cones components corresponds to the
central peak and the peaks on the outside of pulse profiles,
respectively.
Each component of pulse profile is assumed to have a Gaussian shape and can be separated with Gaussian fitting~\citep{WU92,KR94,WU98}.
The cross section structure of the emission beam can be deduced by the average pulse profile~\citep{Oster77,RN83,MR95,QO98,Han01}.
In the magnetosphere of pulsar, the open field line region is divided into two parts by CFL.
One part containing the magnetic axis is called the CG~\citep{RN75} and the other part is the AG~\citep{Qiao03a,Qiao03b,Q04,Q07}.
The locations and shapes of particle acceleration regions (CG and AG) above the polar cap can be determined by calculating the radiation altitude $r$ (the distance from the pulsar center to the radiation point) of each component of pulse profile and the beaming angle $\theta_{\mu}$ (the angle between the radiation direction and the magnetic axis).
\subsection{Component-separation with Gaussian Fitting}
\label{Sect.3.1}
\begin{figure*}
\includegraphics[width=1.45\columnwidth]{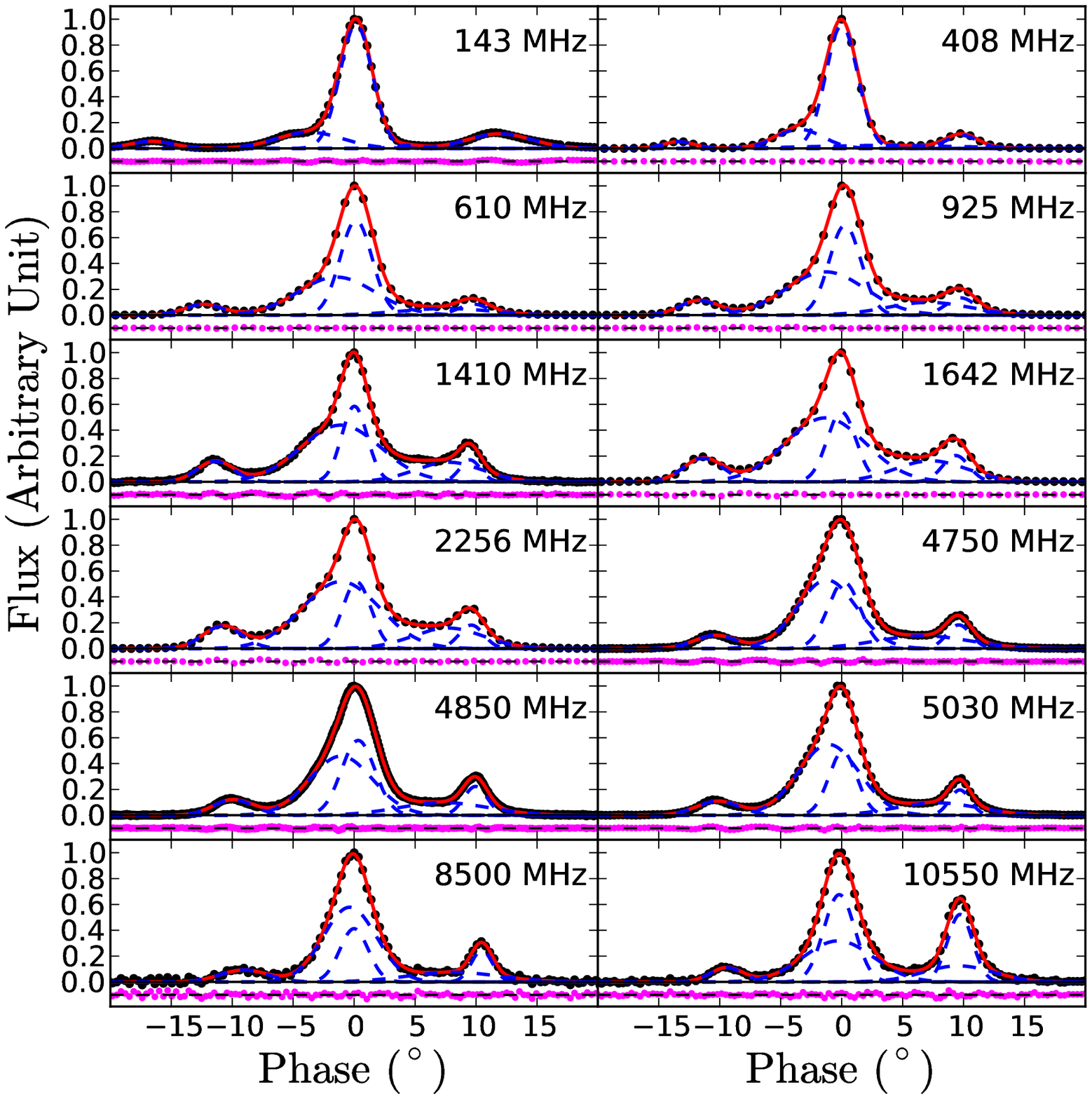}
\caption{Gaussian fitting of PSR B0329$+$54 multi-frequency
profiles in this figure. The black dots are the observation data.
The red solid and blue dash lines are the fitting curves and
single Gaussian respectively. The purplish red dots are the
residuals. The maximum pulse flux is normalized, and the profiles
are aligned at the position of central peak.\label{f1}}
\includegraphics[width=1.45\columnwidth]{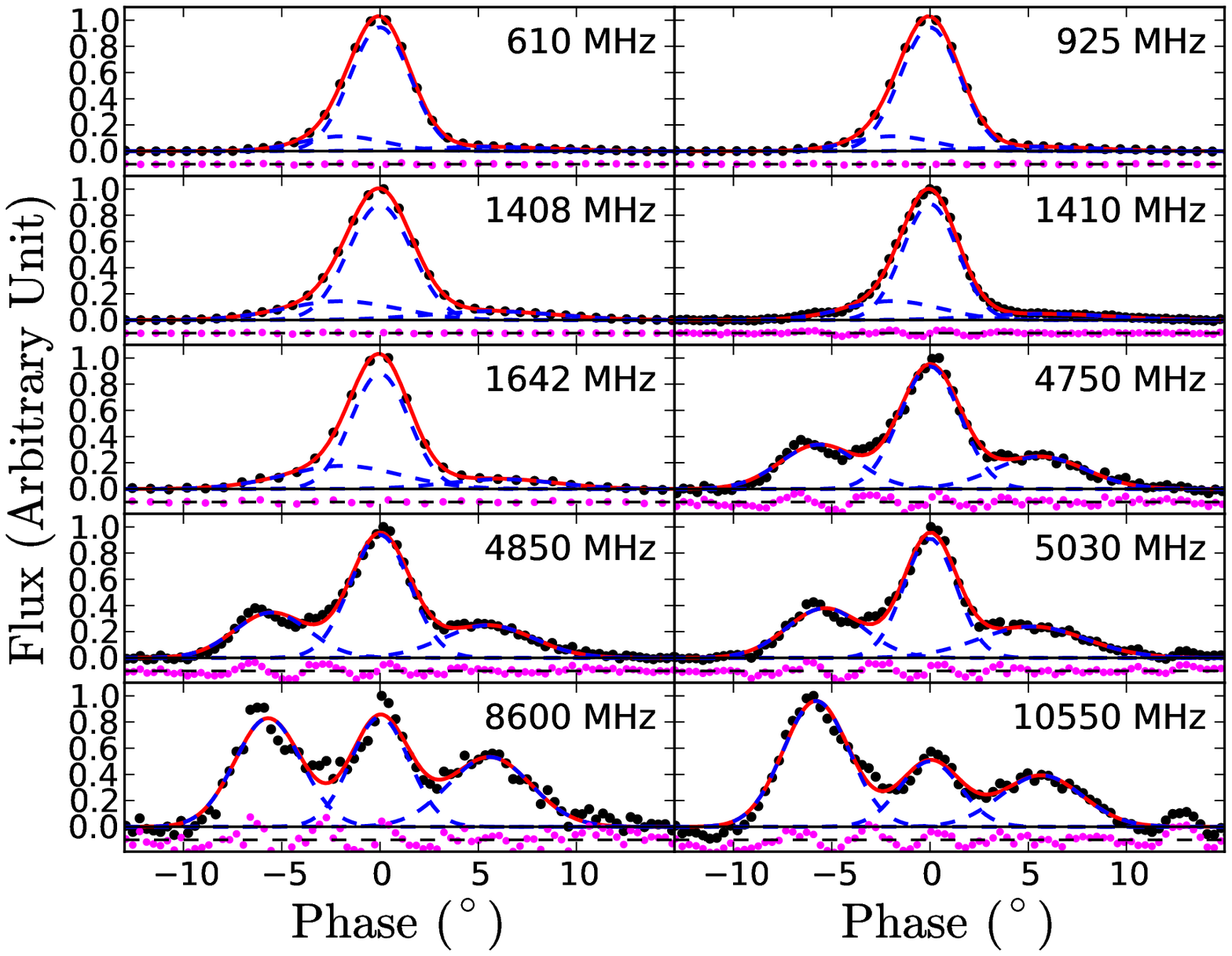}
\caption{As Fig.\,\ref{f1}, but for PSR B1642$-$03.\label{f2}}
\end{figure*}
\begin{table*}
\small
\caption{The Gaussian fitting parameters of PSR B0329$+$54.\label{2}}
\begin{tabular}{ccr@{$\pm$}lr@{$\pm$}lr@{$\pm$}rcccr@{$\pm$}lr@{$\pm$}lr@{$\pm$}r}
\hline
$f$\,(MHz) & $i$ & \multicolumn{2}{c}{$p_i$} & \multicolumn{2}{c}{$I_i$} & \multicolumn{2}{c}{$w_i$} & & $f$\,(MHz) & $i$ & \multicolumn{2}{c}{$p_i$} & \multicolumn{2}{c}{$I_i$} & \multicolumn{2}{c}{$w_i$}\\
\hline
  &   1   &   -16.767   &   0.07   &   0.056   &   0.002   &   1.766   &   0.07     &  &   &   1   &   -11.007   &   0.05   &   0.173   &   0.004   &   1.604   &   0.05     \\
    &   2   &   -3.838   &   0.19   &   0.126   &   0.003   &   2.667   &   0.13     &  &   &   2   &   -1.298   &   0.15   &   0.518   &   0.02   &   3.188   &   0.07     \\
143 &   3   &   \multicolumn{2}{r}{0.0}   &   0.96   &   0.01   &   1.322   &   0.01   & &2256   &   3   &   \multicolumn{2}{r}{0.0}   &   0.516   &   0.02   &   1.162   &   0.03   \\
    &   4   &   4.208   &   0.17   &   0.024   &   0.003   &   1.126   &   0.19   &   & &   4   &   7.492   &   0.44   &   0.159   &   0.01   &   2.896   &   0.26    \\
    &   5   &   11.769   &   0.05   &   0.114   &   0.002   &   2.604   &   0.05    &  &  &   5   &   9.401   &   0.05   &   0.182   &   0.02   &   1.006   &   0.07    \\
\hline
    &   1  &   -13.397   &   0.03   &   0.053   &   0.001   &   1.262   &   0.03   & & &   1   &   -10.542   &   0.05   &   0.103   &   0.003   &   1.708   &   0.05     \\
    &   2   &   -3.47   &   0.06   &   0.148   &   0.003   &   1.906   &   0.05    &  &  &   2   &   -1.409   &   0.18   &   0.531   &   0.04   &   2.423   &   0.06     \\
408 &   3   &   \multicolumn{2}{r}{0.0}   &   0.951   &   0.004   &   1.299   &   0.004   & &4750   &   3   &   \multicolumn{2}{r}{0.0}   &   0.514   &   0.04   &   1.341   &   0.04   \\
    &   4   &   3.395   &   0.62   &   0.023   &   0.001   &   5.788   &   0.31   &  &  &   4   &   6.481   &   0.73   &   0.099   &   0.003   &   3.754   &   0.41    \\
    &   5   &   9.847   &   0.02   &   0.101   &   0.001   &   1.27   &   0.021    & &  &   5   &   9.423   &   0.03   &   0.183   &   0.01   &   0.988   &   0.04    \\
\hline
    &   1   &   -12.53   &   0.03   &   0.084   &   0.001   &   1.57   &   0.03     &  &  &   1   &   -10.25   &   0.02   &   0.121   &   0.001   &   1.629   &   0.02     \\
    &   2   &   -1.602   &   0.17   &   0.295   &   0.01   &   3.169   &   0.07     &   & &   2   &   -1.403   &   0.07   &   0.458   &   0.01   &   2.586   &   0.02     \\
610 &  3   &   \multicolumn{2}{r}{0.0}   &   0.733   &   0.01   &   1.308   &   0.01 & &4850   &   3   &   \multicolumn{2}{r}{0.0}   &   0.579   &   0.01   &   1.372   &   0.01   \\
    &   4  &   7.377   &   0.98   &   0.052   &   0.003   &   3.651   &   0.58   &  & &   4   &   6.987   &   0.24   &   0.099   &   0.001   &   3.6   &   0.15    \\
    &   5   &   9.606   &   0.04   &   0.086   &   0.01   &   1.209   &   0.07    &  &  &   5   &   9.617   &   0.01   &   0.223   &   0.003   &   0.959   &   0.01    \\
\hline
    &   1   &   -11.966   &   0.04   &   0.114   &   0.002   &   1.581   &   0.04     &  & &   1   &   -10.262   &   0.05   &   0.109   &   0.002   &   1.768   &   0.05     \\
    &   2   &   -1.586   &   0.33   &   0.335   &   0.02   &   3.267   &   0.14    &  & &   2   &   -1.156   &   0.06   &   0.548   &   0.01   &   2.505   &   0.03     \\
925 &   3   &   \multicolumn{2}{r}{0.0}   &   0.695   &   0.02   &   1.301   &   0.02   & &5030   &   3  &   \multicolumn{2}{r}{0.0}   &   0.492   &   0.01   &   1.149   &   0.02   \\
    &   4  &   6.84   &   1.06   &   0.099   &   0.01   &   3.37   &   0.50    & & &   4   &   7.734   &   0.26   &   0.097   &   0.004   &   3.35   &   0.20    \\
    &   5   &   9.486   &   0.04   &   0.134   &   0.01   &   1.208   &   0.07    &  & &   5   &   9.6   &   0.02   &   0.196   &   0.01   &   0.847   &   0.03    \\
\hline
    &   1   &   -11.427   &   0.04   &   0.159   &   0.004   &   1.479   &   0.04     & & &   1   &   -9.037   &   0.16   &   0.093   &   0.01   &   1.942   &   0.18     \\
    &   2  &   -1.123   &   0.09   &   0.439   &   0.01   &   3.281   &   0.05     &  & &   2   &   -0.352   &   0.08   &   0.578   &   0.09   &   2.184   &   0.12     \\
1410   &   3   &   \multicolumn{2}{r}{0.0}   &   0.585   &   0.01   &   1.055   &   0.02   & & 8500   &   3   &   \multicolumn{2}{r}{0.0}   &   0.413   &   0.09   &   1.167   &   0.10   \\
    &   4   &   7.719   &   0.23   &   0.152   &   0.01   &   2.751   &   0.15   & & &   4   &   8.132   &   0.74   &   0.076   &   0.01   &   3.717   &   0.58    \\
    &  5   &   9.482   &   0.03   &   0.17   &   0.01   &   0.847   &   0.05   & & &   5   &   10.483   &   0.04   &   0.244   &   0.01   &   0.858   &   0.05    \\
\hline
    &   1   &   -11.365   &   0.04   &   0.182   &   0.004   &   1.596   &   0.04    &  & &   1   &   -9.255   &   0.08   &   0.111   &   0.01   &   1.338   &   0.08     \\
    &   2   &   -1.326   &   0.18   &   0.495   &   0.02   &   3.291   &   0.09     &  &&   2   &   -0.098   &   0.07   &   0.318   &   0.04   &   2.861   &   0.16     \\
1642   &  3   &   \multicolumn{2}{r}{0.0}   &   0.541   &   0.02   &   1.18   &   0.03  & &10550    &   3   &   \multicolumn{2}{r}{0.0}   &   0.676   &   0.04   &   1.273   &   0.04   \\
    &   4  &   7.338   &   0.52   &   0.163   &   0.01   &   2.962   &   0.29    &  & &   4   &   9.765   &   0.18   &   0.125   &   0.02   &   2.86   &   0.28    \\
    &   5   &   9.375   &   0.04   &   0.204   &   0.02   &   1.08   &   0.07    &  & &   5   &   9.899   &   0.02   &   0.522   &   0.02   &   1.035   &   0.03    \\
\hline
\end{tabular}

\footnotesize{Note: $f$ is the observed frequency, $I_i$, $p_i$,
$w_i$ are the intensity, peak position, standard deviation of the
$i$th Gaussian component, respectively.} \caption{The Gaussian
fitting parameters of PSR B1643$-$03.\label{3}}
\begin{tabular}{ccr@{$\pm$}lr@{$\pm$}lr@{$\pm$}rcccr@{$\pm$}lr@{$\pm$}lr@{$\pm$}r}
\hline
$f$\,(MHz) & $i$ & \multicolumn{2}{c}{$p_i$} & \multicolumn{2}{c}{$I_i$} & \multicolumn{2}{c}{$w_i$} & & $f$\,(MHz) & $i$ & \multicolumn{2}{c}{$p_i$} & \multicolumn{2}{c}{$I_i$} & \multicolumn{2}{c}{$w_i$}\\
\hline
   &   1   &   -1.895   &   2.52   &   0.114   &   0.12   &   2.136   &   0.73    & &    &   1   &   -5.641   &   0.08   &   0.34   &   0.01   &   1.973   &   0.09     \\
610    &   2   &   \multicolumn{2}{r}{0.0}   &   0.946   &   0.17   &   1.523   &   0.05   & & 4750    &   2   &   \multicolumn{2}{r}{0.0}   &   0.937   &   0.01   &   1.529   &   0.04    \\
   &   3   &   5.427   &   0.76   &   0.034   &   0.003   &   3.069   &   0.48    & &  &   3   &   5.589   &   0.15   &   0.247   &   0.01   &   2.314   &   0.15    \\
\hline
   &   1   &   -1.894   &   2.35   &   0.114   &   0.11   &   2.136   &   0.68    & &   &   1  &   -5.526   &   0.07   &   0.346   &   0.01   &   1.877   &   0.07     \\
925    &   2   &   \multicolumn{2}{r}{0.0}   &   0.946   &   0.16   &   1.523   &   0.05  & &4850   &   2   &   \multicolumn{2}{r}{0.0}   &   0.939   &   0.01   &   1.506   &   0.03    \\
   &  3   &   5.427   &   0.71   &   0.034   &   0.003   &   3.069   &   0.45  & & &   3  &   5.497   &   0.12   &   0.25   &   0.01   &   2.281   &   0.13    \\
\hline
   &  1   &   -1.991   &   0.76   &   0.144   &   0.03   &   2.886   &   0.25    & &  &   1  &   -5.344   &   0.07   &   0.38   &   0.01   &   1.942   &   0.08     \\
1408   &   2   &   \multicolumn{2}{r}{0.0}   &   0.877   &   0.04   &   1.619   &   0.02  & &5030   &   2   &   \multicolumn{2}{r}{0.0}   &   0.912   &   0.02   &   1.329   &   0.03    \\
   &   3   &   5.664   &   0.65   &   0.067   &   0.01   &   3.249   &   0.31   & & &   3   &   5.202   &   0.20   &   0.239   &   0.01   &   2.709   &   0.20    \\
\hline
   &   1   &   -1.908   &   0.99   &   0.145   &   0.04   &   2.437   &   0.33     & &  &   1   &   -5.697   &   0.08   &   0.829   &   0.03   &   1.693   &   0.08     \\
1410   &  2   &   \multicolumn{2}{r}{0.0}   &   0.886   &   0.07   &   1.397   &   0.04  & &8600   &   2   &   \multicolumn{2}{r}{0.0}   &   0.839   &   0.03   &   1.482   &   0.10   \\
   &   3   &   5.914   &   0.88   &   0.048   &   0.004   &   3.191   &   0.57   & &  &   3   &   5.578   &   0.16   &   0.533   &   0.03   &   2.102   &   0.17    \\
\hline
   &   1   &   -1.896   &   1.18   &   0.177   &   0.05   &   2.959   &   0.43    & &  &   1   &   -5.798   &   0.06   &   0.961   &   0.03   &   1.676   &   0.07     \\
1642   &   2   &   \multicolumn{2}{r}{0.0}   &   0.881   &   0.07   &   1.471   &   0.05    & &10550  &   2   &   \multicolumn{2}{r}{0.0}   &   0.5   &   0.03   &   1.542   &   0.16    \\
   &   3   &   6.201   &   1.06   &   0.074   &   0.01   &   2.778   &   0.61   & & &   3   &   5.665   &   0.20   &   0.393   &   0.02   &   2.069   &   0.21    \\
\hline
\end{tabular}

\footnotesize{Note: with the same parameters as Table\,\ref{2}}
\end{table*}

\begin{figure*}
\centering
\includegraphics[width=0.90\columnwidth]{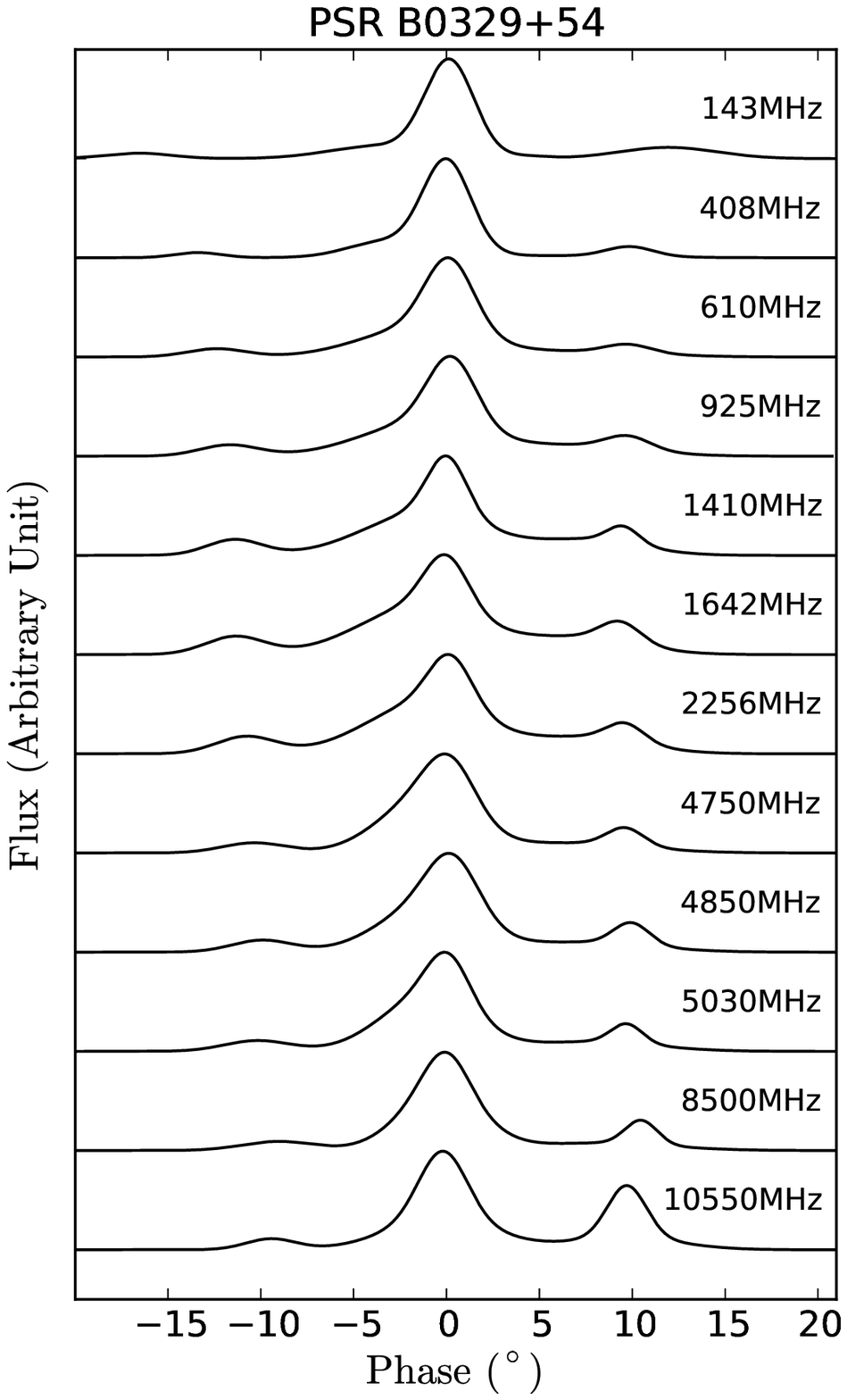}
\includegraphics[width=0.90\columnwidth]{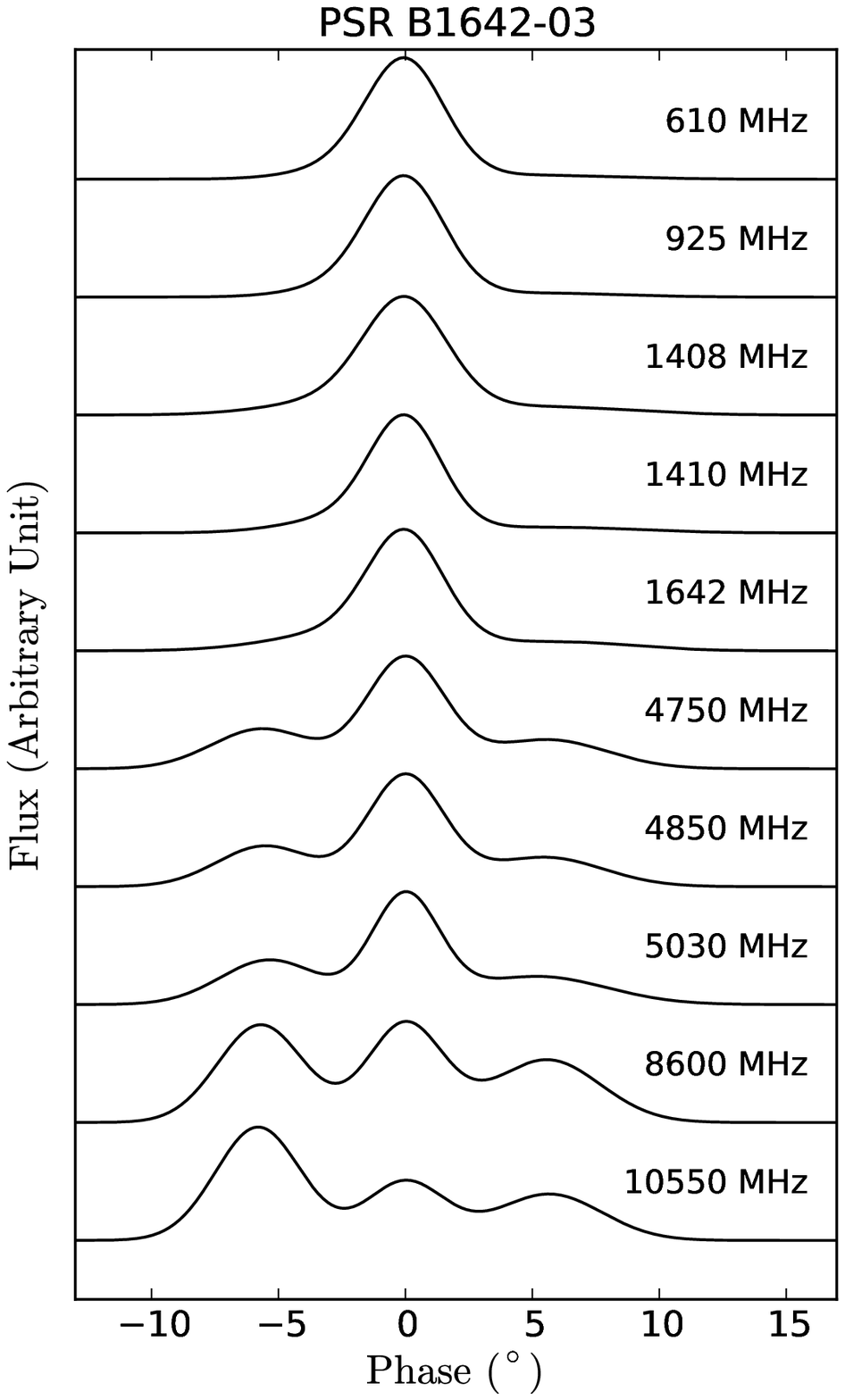}
\caption{The fitting curves evolution versus the observing frequency.\label{f3}}
\end{figure*}
\begin{figure*}
\centering
\includegraphics[width=1.70\columnwidth]{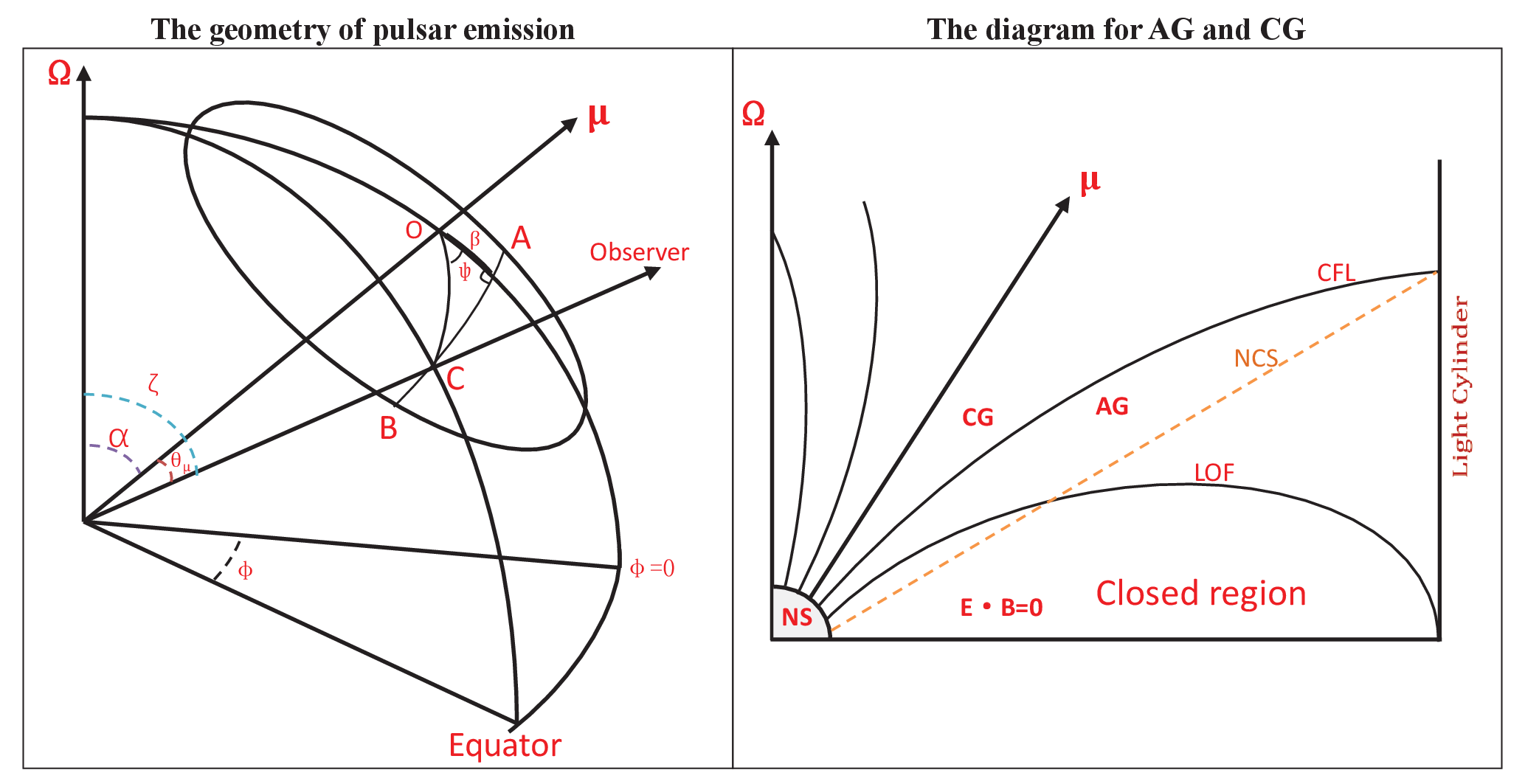}
\caption{Left: the geometry of pulsar emission~\citep{ZG07}.
$\Omega$ and $\mu$ are the rotation axis and the magnetic axis.
$\zeta$, $\alpha$, $\beta = \zeta - \alpha$ and $\theta_{\mu}$ are
the angle of the observer's line of sight to the rotation axis,
the magnetic inclination angle, the impact angle and the angle of
radiation direction to the magnetic axis, respectively. $\phi$
represents the azimuth angle around the rotation axis, $\psi$ is
the angle between the magnetic field plane and the $\Omega$-$\mu$
plane. C is a point that we observe the radiation when the line of
sight sweeps the emission beam from $A$ to $B$. Right: a simple
diagram for the AG and the CG in the magnetosphere of neutron star
(NS)~\citep{Q07}. ``NCS" represents the null change surface where
$\Omega \cdot B = 0$. ``CFL" and ``LOF" are the CFL and the last
open field line respectively.\label{f4}}
\end{figure*}

\begin{figure*}
\includegraphics[width=1.99\columnwidth]{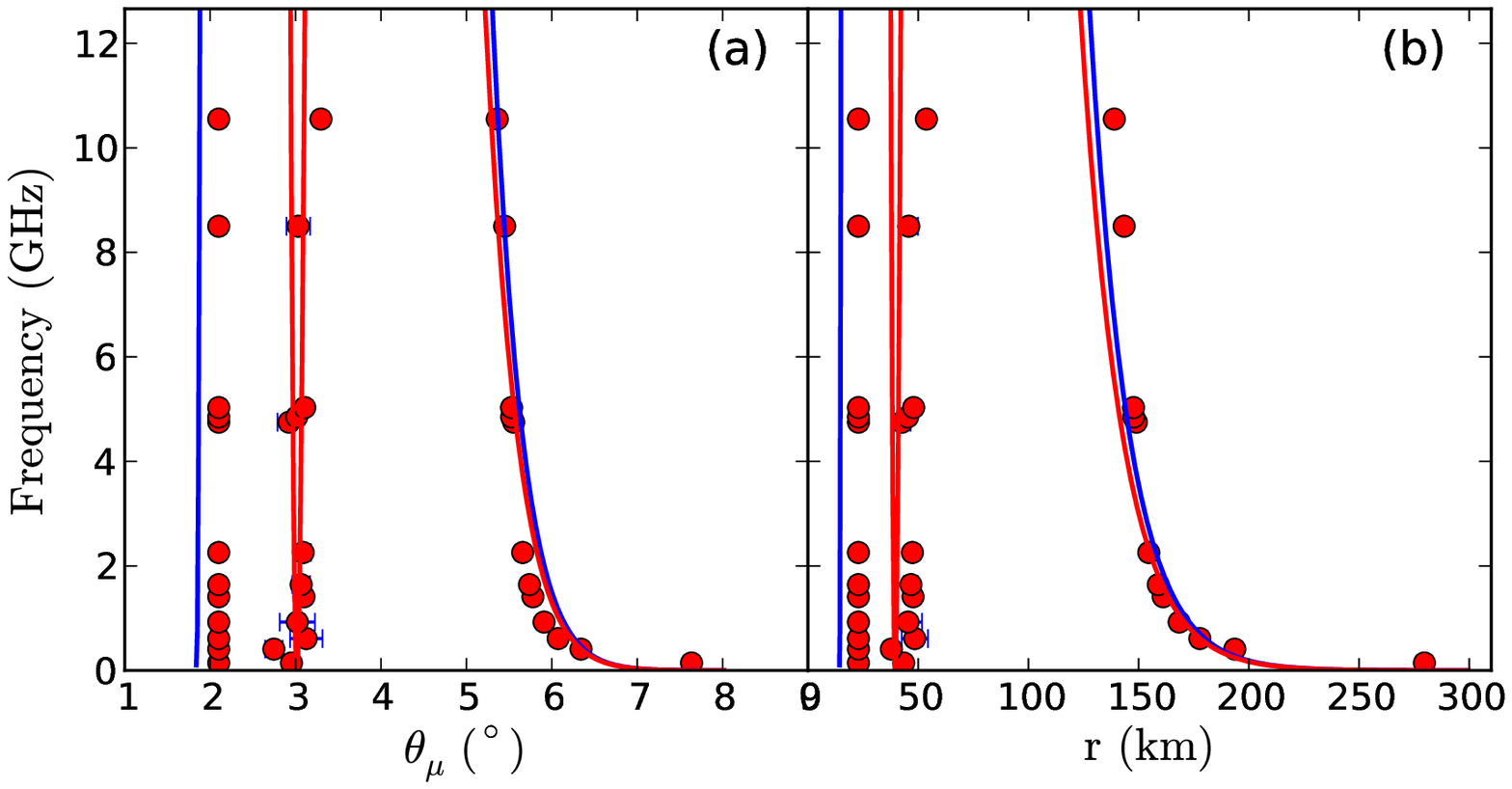}
\caption{The simulating results of beam-frequency evolution of PSR B0329$+$54 in the ICS model.
The figure\,(a) and (b) present the evolutions of beaming angle
$\theta_{\mu}$ and the emission altitudes versus frequency
respectively.
The red dots are the emission altitudes.
The blue and red solid lines are the ICS fitting curves.
The values of $\gamma_0$ and $k$ are set to $2 \times 10^5$ and
0.329 respectively.\label{f5}}
\includegraphics[width=1.99\columnwidth]{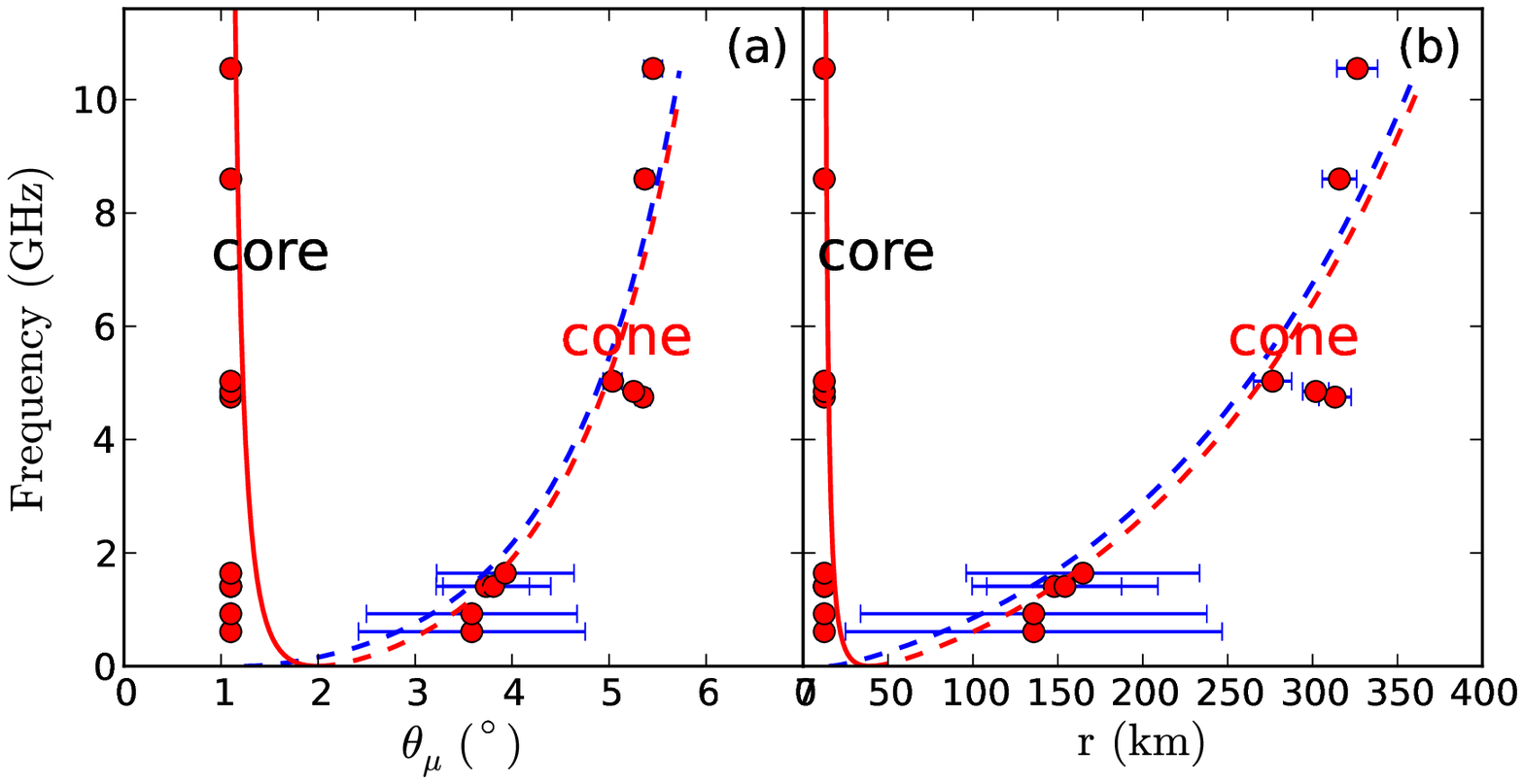}
\caption{As Fig.\,\ref{f5}, but for PSR B1642$-$03. $\gamma_0=1.6\times 10^3$, $k=-0.02$.
The fitting curves are plotted with solid curves for core
component and dashed curves for cone component.\label{f6}}
\end{figure*}
\begin{figure}
\centering
\includegraphics[width=0.99\columnwidth]{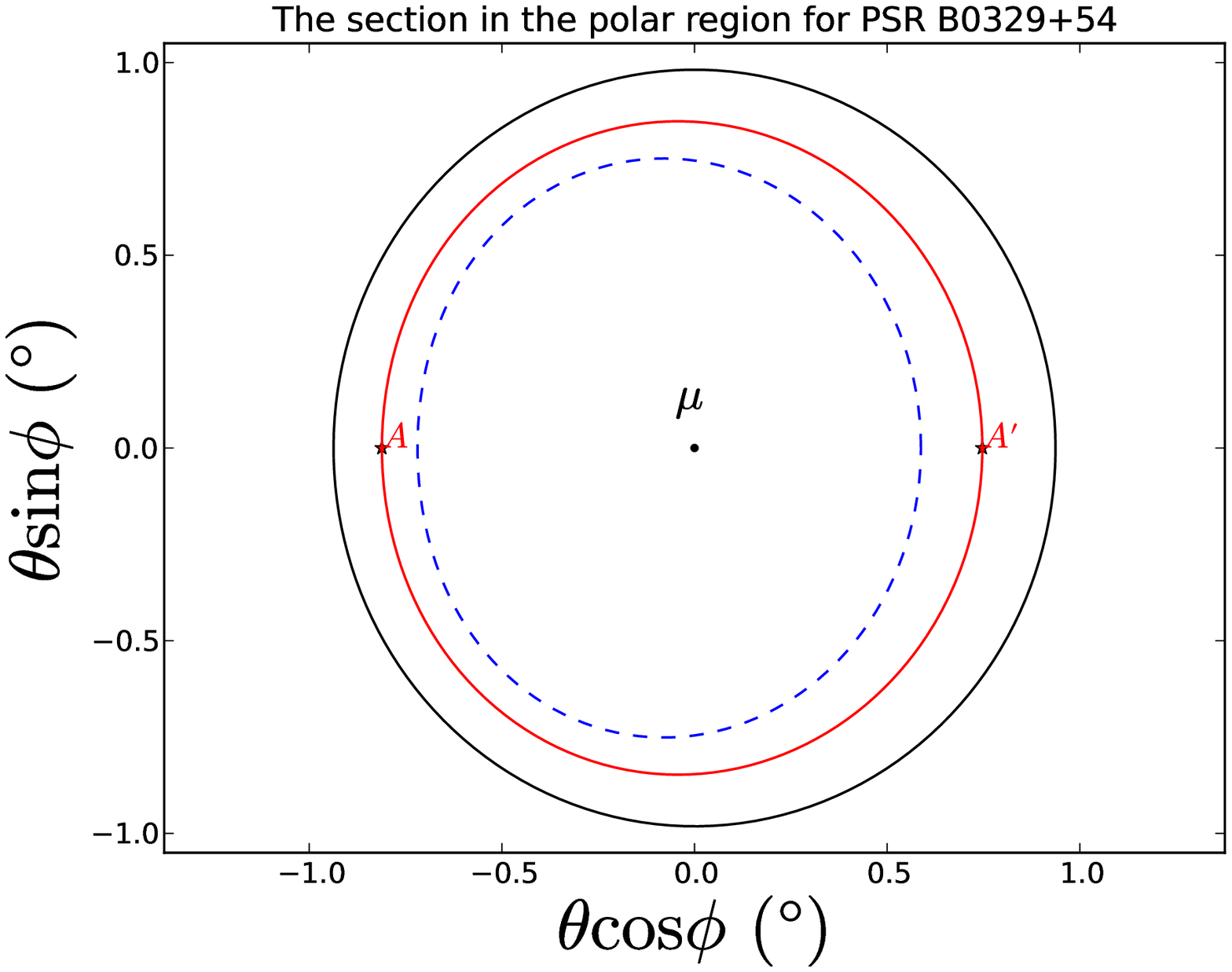}
\includegraphics[width=0.99\columnwidth]{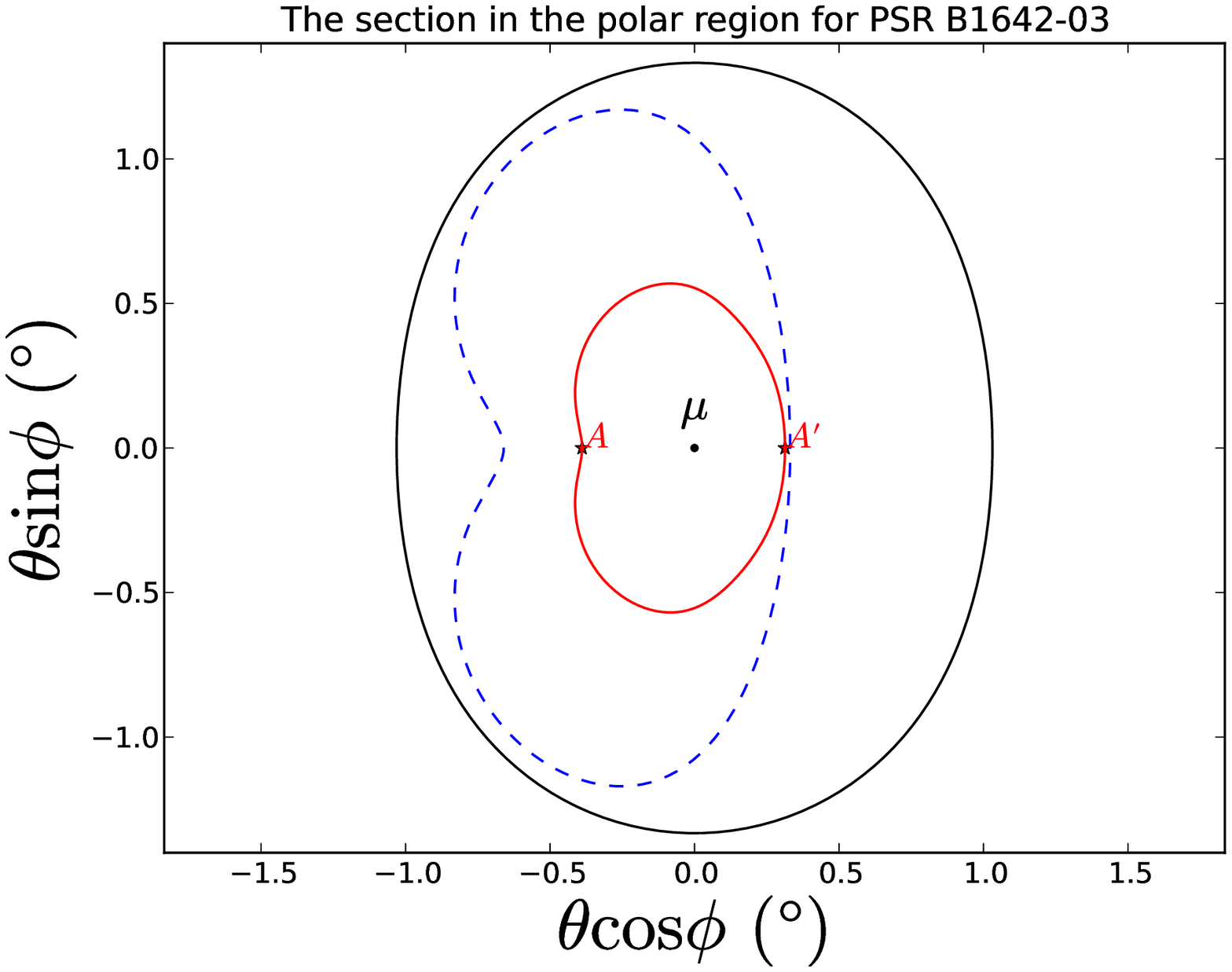}
\caption{The black solid and blue dashed lines are the LOF and CFL
respectively.
The region between blue dashed line and black solid line is AG, the region surrounded by blue dashed line is the CG.
The red solid line is the magnetic field line where the radiation comes from.
The red solid line is also the section of all the sparking points
in the AG or the CG.
The red star dots $A$ and $A'$ are the sparking points used in our calculation.
For PSRs B0329$+$54 and B1642$-$03, the values of $\alpha$ are
$30^\circ.0$ and $68^\circ.2$ respectively.\label{f7}}
\end{figure}

Usually, the pulse profile is interpreted as the composition of
different radiation components from different altitudes.
Each component of the pulse profiles is assumed to have a Gauss
shape and can be expressed by Gaussian functions.
Here the multi-gaussian fitting method~\citep{WU92,KR94,WU98} is
used to separate the different components of the pulse profiles.
The Gaussian functions used in fitting can be written as:
\begin{equation}
\label{eq1}
y = \sum\limits_{i=1}^N {{I_i}\exp \left[-\frac{(x-p_i)^2}{2{w_i}^2}\right]},
\end{equation}
where $I_i$, $p_i$ and $w_i$ are intensity, position of peak and
standard deviation of $i$th Gaussian respectively.
However, the residual exists between the results of model fitting
and the observed data.
In order to make the fitting results coincide better with the
observed data.
fitting parameters $I_i$, $p_i$ and $w_i$.
For a good fitting, the residual curve of the on-pulse part shall
be noiselike, and as similar as possible to that of the off-pulse
part.
The final fitting parameters of various components of the pulse profiles of PSRs B0329$+$54 and B1642$-$03 are shown in Table\,\ref{2} and \ref{3},
and the corresponding fitting curves are shown in Fig.\,\ref{f1} and \ref{f2}.
Besides, the evolution of pulse profiles versus frequencies are shown in Fig.\,\ref{f3}.

For PSR B0329$+$54, the integrated profiles can be separated into five~\citep{KR94} or even nine components~\citep{Gang01}.
In this work, however, we fit each observational pulse profile with 5 Gaussian components.
For PSR B1642$-$03, we fit each pulse profile with 3 Gaussian components to obtain satisfied residual.
As the fitting results in Fig.\,\ref{f1}-\ref{f3}, for PSR
B0329$+$54 the integrated pulse profiles show narrower profiles at
higher frequencies.
Each integrated pulse profile of PSR B1642$-$03 has only a core
and one inner cone component, and the pulse profiles at higher
frequencies are wider than those at lower frequencies.
These results coincide with the prediction by~\cite{QO98,QOL1},
 i.e. the slow-rotation pulsars prefer to be ``double-cones pulsars", while the fast-rotation pulsars tend to show triple profiles.

\subsection{Beam-frequency Evolution}
\label{Sect.3.2} We calculate geometrically the different
radiation altitudes of components in various frequencies, and
simulate the evolutions of radiation altitudes and beaming angles
versus frequency in the ICS model.
First, the magnetic field of pulsar is assumed to be dipole with
the last open field line\,(LOF) tangent to the light
cylinder\,(see~\citep{ZG07} for detailed equations).
Secondly, the radiation is assumed to come from the open magnetic field line with the radiation direction tangent to that line.
The radius of the magnetic field line
where the radiation comes from is a specific times that of LOF.
Thirdly, the radiation altitudes $r = \eta^{-2} R_\text{e} sin^2{\theta}$ of radiation components at a certain frequency can be determined with the parameters in Table\,\ref{2}-\,\ref{3} and the inclination angle $\alpha$ and the impact angle $\beta$ in Fig.\,\ref{f4}.
$R_\text{e}$ is the maximum radius of LOF.
Obviously, for $\eta=1$, the radiation is generated from LOF.
$\theta$ is the polar angle between the magnetic axis and the radiation altitude $r$.
In our calculation, the values of $\alpha = 30^{\circ}$ and $\beta = 2^{\circ}.1$~\citep{RN93} for PSR B0329$+$54,
 and $\alpha = 68^{\circ}.2$, $\beta =1^{\circ}.1$~\citep{Lyne88} for PSR B1642$-$03 are used respectively.

For the ICS process, the frequency $\omega'$ of the radiation photon (with the
energy of $\hbar \omega'$) can be determined by the following equation (see~\citealt{QO98}):
\begin{equation}
\label{eq2}
\omega' = 2 {\gamma ^2}{\omega _0}(1 - \eta \cos {\theta _i}),
\end{equation}
where $\theta_i$ is the angle between the outgoing direction\,(initial direction of photons moving from the sparking point,
see sparking point $A$ in Fig.1 in~\citealt{QO98}) and incoming direction (radiation direction) of photons; $\omega _0 = 2\pi\times10^{6}\,\rm{rad/s}$~\citep{LU16} is the initial angular frequency of the low frequency wave; $\gamma$ is the Lorentz factor of particle with the value $\gg 1$; $\eta=\frac{v}{c}\approx 1$.
$\theta_{i}$ can be determined by the following equation~\citep{QO98}:
\begin{equation}
\label{eq3}
\cos {\theta_i} = \frac{{2\cos \theta  + \left( {1 - 3{{\cos }^2}\theta } \right)\frac{R}{r}}}{{{{\left( {1 + 3{{\cos }^2}\theta } \right)}^{1/2}}\left( {1 - 2\frac{R}{r}\cos \theta  + \frac{R}{r}} \right)}},
\end{equation}
where $R$ is the radius of pulsar.
In our calculation, we consider a case that the low frequency wave is produced by the electron oscillating at the sparking points $A$ or $A'$ (\,see Fig.\,\ref{f7}\,) in the gap.
The high energy particles will lose energy owing to be scattered
by the thermal photons and low frequency waves when the particles
move along with the field lines. The lorentz factor of the
particle is assumed to vary with the exponential form: $\gamma =
{\gamma _0}\exp\left(-k\frac{r-R}{R}\right)$, where $\gamma_0$ is
the initial lorentz factor, $k$ is a factor for the energy loss of
particles. The locations of the sparking points and the values of
$\gamma_0$ and $k$ can be obtained by the ICS model simulating the
evolutions of radiation altitudes and beaming angles versus the
radiation frequency.

The simulative results of the ICS model are shown in~Fig.\,\ref{f5}\,(PSR B0329$+$54) and Fig.\,\ref{f6}\,(PSR B1642$-$03), where the red and blue solid (\,dashed\,) lines represent the theoretical curves of ICS model that the observed waves outgoing from the points $A$ and $A'$ (see~Fig.\,\ref{f7}), respectively.
Fig.\,\ref{f5}(a) and Fig.\,\ref{f6}(a) present the beaming angle evolutes with the radiation frequency.
While Fig.\,\ref{f5}(b) and Fig.\,\ref{f6}(b) show the radiation altitude evolutes with the radiation frequency.
By the ICS model simulating the evolutions of radiation altitude and beaming angle versus the radiation frequency, the parameter values of $\gamma_0 \sim 2 \times 10^5$ and $k\sim 0.329$ for PSR B0329$+$54, and $\gamma_0 \sim 1.6\times 10^3$ and $k\sim -0.02$ for PSR B1642$-$03 are obtained, respectively.
Meanwhile, the field lines where the radiation comes from and the sparking points $A$ and $A'$ are also determined by the simulation.
The foots of the field lines above the polar cap are located at
$0.98\,R_\text{p}$ for PSR B0329$+$54 and $0.46\,R_\text{p}$ for
PSR B1642$-$03 respectively.
The locations of sparking points above the polar cap are
$0.98\,R_\text{p}$ for PSR B0329$+$54 and $0.46\,R_\text{p}$ for
PSR B1642$-$03 respectively, where
$R_\text{p}=R^{1.5}\Omega^{0.5}c^{-0.5}$ represents the polar cap
radius is defined by the foot of LOF ($R$, $\Omega$ and $c$ are
the pulsar radius, rotation angular velocity and light speed,
respectively~\citealt{RN75}). In~Fig.\,\ref{f5}-\ref{f6}, the
theoretical curve of the ICS model shows a ``down-up-downward"
trend, which is consistent with the conclusion
of~\citep{QO98,ZG07}.
For each profile, the radiation altitudes of the inner cone components are lower than those of the outer cone components.
The radiation altitudes of the outer components decrease with the radiation frequency increasing.
While, the radiation altitudes of inner cone components increase with the frequency increasing.
\subsection{Radiation Region}
\label{Sect.3.3}
In an oblique rotator with a dipolar magnetic field, the shape of gap is a function of the pulsar period and inclination angle $\alpha$~\citep{Q04,Wang06,Lee06a}.
The CG and AG in the open magnetic field region are shown in Fig.\ref{f4}.
~\citealt{Q04} points out that the sparking may take place in
these two gaps, which will lead to pairs production and generate
the secondary particles.
These particles would be accelerated in the CG or AG to produce radiation through the ICS process.
The radiation region of the pulsar can be determined conclusively
if the sparking point and the magnetic field line where the
radiation comes from are determined (the detailed calculation
methods are given by~\citet{Q04,Wang06,Lee06a}).

We try to determine the radio radiation regions of PSRs B0329$+$54
and B1642$-$03 by using the simulative results
in\,\S\ref{Sect.3.2}. The shapes and widths of the radiation
regions above the polar cap are shown in~Fig.\,\ref{f7}.
For PSR B0329$+$54, the section\,(\,the red solid line\,) of the
foots of field lines where the radiation comes from and the
sparking points $A$ and $A'$ above the polar cap are located at
the region between those of the LOF\,(the black solid line) and
CFL\,(the blue dashed line). We also obtain a simulation of
beam-frequency, though the simulation is not optimal considering
this case that the radiation is generated from the CG.
Therefore, the CG and AG can be responsible for the radiation of
PSR B0329$+$54.
For PSR B1642$-$03, however, the section\,(\,the red solid line\,)
of the foots of field lines where the radiation comes from and the
sparking points above the polar cap are located at the region
surrounded by the CFL, which means that the radiation of PSR
B1642$-$03 is generated mainly from the CG.
\section{CONCLUSIONS AND DISCUSSIONS}
\label{Sect.4} With the multi-Gaussian function, we separate the
radiation components of the multi-band radio pulse profiles of
PSRs B0329$+$54 and B1642$-$03, and calculate the radiation
altitude of each radiation component (\,$r=20\sim280$\,km for PSR
B0329$+$54 and $r=10\sim330$\,km for PSR B1642$-$03\,).
Then we simulate further the evolutions of radiation altitudes and
beaming angles with the ICS model and obtain the Lorenz factors
(\,$\gamma_0\simeq 10^5$ for PSR B0329$+$54 and $\gamma_0\simeq
10^3$ for PSR B1642$-$03\,) and energy loss factors of high energy
particles\,(\,$k=0.329$ for PSR B0329$+$54 and $k=-0.02$ for PSR
B1642$-$03\,).
Finally, we get the radio radiation regions of the two sources.
\begin{enumerate}[(1)]
\item It is suspected that the radio emission beam of pulsars may
include core and multi-cone
components~\citep{RN83,RN93,Lyne88,WU92,WU98,MR95,ZG07}. Now, the
pulse profiles of PSRs B0329$+$54 and B1642$-$03 show obvious
multi-peak, which can be described by multi-Gaussian functions
(\,5 Gaussians for PSR B0329$+$54 and 3 Gaussians for PSR
B1642$-$03, see Fig.\ref{f1} and Fig.\ref{f2} for details\,),
which supports the scenario of core and cone components.
However, the formation and radiation mechanism of multi-components are still disputable.
We find that the multi-components of the two sources can be described well by the ICS model,
which implies that the core and cones components may be generated by the same radiation mechanism.
This is consistent with the conclusion of~\citet{Lyne88}.

\item In most cases the integrated pulse profiles of pulsars are extremely stable.
However, in case of a moding pulsar~\citep{Backer70,Helfand75,Wang07,Kijak98,van Leeuwen02,Redman05},
    during an observation, if a mode change occurs, the observed profile will be different from the profile of any individual mode.
    Furthermore, when moding occurs, the observed pulse profile will be no longer stable because the relative contribution of each mode is different.
    For PSR 0329$+$54, the mode switching of the pulse profile is observed~\citep{Lyne71,Bart82,Hesse73,Izvekova94,Kuzmin96,Sule02,Chen11}.
Does this mode changing influence our results?~\citet{Kuzmin96}
found that for PSR 0329$+$54, ``When the mode changes, the
component structure (\,the number of the components and their
positions in phase\,) does not change;
only the relative intensity of the components varies".
We also get the same conclusion by analyzing some mode changing data of PSR B0329$+$54.
Therefore the mode switching phenomenon does not influence our
result. In future, we plan to use large-aperture telescopes (e.g.,
FAST) for long-term monitoring of some pulsars including these two
pulsars and study the pulse profiles evolution with the new data
and the ICS model thus to verify the conclusions presented in this
paper.
\item It is very important to determine the radiation locations by
analyzing the observed multi-frequency pulse profiles for
understanding how the radiation is generated in pulsar
magnetosphere and constraining the pulsar radiation models. For
PSRs B0329$+$54 and B1642$-$03, the sparking and related radiation
can take place in both CG and AG. Can we observe the radiation of
these two pulsars from both of CG and AG?
The answer depends on the inclination angle $\alpha$ and the
impact angle $\beta$ (or viewing angle $\zeta=\alpha + \beta$).
From the observed values of $\alpha $ and $\beta$~\citep{Lyne88,RN93}, under the dipole magnetic field assumption,
 radiation regions are calculated geometrically.
It is found that the CG or AG can be responsible for the radio
radiation of PSR B0329$+$54.
Whereas, the radiation is likely to be generated just in the CG for PSR B1642$-$03.
Besides the particle acceleration and radiation locations, another question is: what shapes of pulse profiles can be observed?
The shapes of pulse profiles change with frequencies.
The beam-frequency evolution presents a constraint to all theoretical models.
The ICS model has two typical ``beam-frequency map" (The observing frequency is plotted versus the beaming angle or altitudes of radiation points)~\citep{QO98,QOL1}.
The ``beam-frequency map" of PSRs B0329$+$54 and B1642$-$03 (Fig.\,\ref{f5} and Fig.\,\ref{f6} in this paper) show these two types respectively.
Especially the ``beam-frequency map" of PSR B1642$-$03 shows that
the pulse profiles become wider at higher frequencies, which
challenges other theoretical models.

\item The Lorentz factors we used for simulating the observed data are $\sim 10^5$ for PSR B0329$+$54, and $\sim 10^3$ for PSR B1642$-$03 (\,see Fig.\ref{f5} and Fig.\ref{f6}\,).
These Lorentz factors are larger than those (\,$\gamma \simeq \ 800$\,) of secondary particles and lower than the energy of the primary particles $\gamma_0 \simeq \ 10^6$ given by~\citet{RN75}.
In the CR model~\citep{RN75}, pair production needs the Lorentz factors of primary particles as large as $\ 10^6$.
As discussed by~\cite{ZG07}, not all pairs are produced at the
bottom of the gap, therefore not all primary particles can gain
such a large Lorentz factor.
When the pairs are generated outside the gap, the Lorentz factor
shall be $\lesssim 10^6$.
This means the Lorenz factor $\gamma_0$ used in this paper is reasonable.
\end{enumerate}
\section*{Acknowledgements}
\label{Sect.6}
This work was supported by the National Basic Research Program of China (Grant No.2012CB821800),
the International Partnership Program, the Strategic Priority Research Program and the Strategic Priority Research Program ``The
Emergence of Cosmological Structures" of CAS (Grant Nos.114A11KYSB20160008, XDB23000000 and XDB09000000),
the NSFC (Grant Nos. 11565010, 11373011, 11225314, 11033008, 11165005, 11173046, 11403073, 11303069, 11173041), the science and technology innovation talent team (Grant No. (2015)4015), the Training Program for Excellent Young Talents (Grant No. 2011-29), the High level Creative Talents (Grant No.(2016)-4008) and the Innovation Team Foundation of the Education Department (Grant No. [2014]35) of Guizhou Province, the Natural Science Foundation of Shanghai No. 13ZR1464500, the Scientific Program of Shanghai Municipality (08DZ1160100), the Knowledge Innovation Program of CAS (grant No. KJCX1-YW-18), the Strategic Priority Research Program The Emergence of Cosmological Structures of CAS (grant No. XDB09000000), Doctoral Starting up Foundation of Guizhou Normal University 2014, Science and Technology Foundation of Guizhou Province (Grant Nos. J[2015]2113 and LH[2016]7226) and the Postgraduate Innovation Foundation of Guizhou normal university (Grant No. 201527).
This work is also supported partially by the fund of ``Special
Demonstration of Space Science".

We would like to thank the Yunnan Astronomical Observatory, CAS
$\&$ Shanghai Astronomical Observatory, CAS, and the European
pulsar network database for providing observation condition and
valuable data resources.

%
%
\bsp    
\label{lastpage}
\end{document}